\documentclass{elsart} 
\usepackage{epsfig}

\begin{document}

\begin{frontmatter}

\title{Condensation in the Backgammon model}
\author[Amsterdam]{Piotr Bialas\thanksref{pperm}},
\author[Bielefeld]{Zdzislaw Burda\thanksref{zperm}},
\author[Edinburgh]{Des Johnston}

\address[Amsterdam]{
{Universiteit van Amsterdam, Instituut voor Theoretische
Fysica,}\\ 
{ Valckenierstraat 65, 1018 XE Amsterdam, The
Netherlands}} 
\thanks[pperm]{Permanent address: Institute of Comp. Science,
Jagellonian University,\\ ul. Nawojki 11, 30-072 Krak\'ow, Poland}
\address[Bielefeld]{Fakult\"{a}t f\"{u}r Physik, Universit\"{a}t
Bielefeld,\\ Postfach 10 01 31, Bielefeld 33501, Germany}
\thanks[zperm]{Permanent address: Institute of Physics, 
Jagellonian University, ul. Reymonta 4, 30-059 Krak\'{o}w, Poland}
\address[Edinburgh]{Dept. of Mathematics, Heriot--Watt University,\\ Edinburgh EH14 4AS, Scotland}

\begin{abstract}
We analyse the properties of a very simple ``balls--in--boxes'' model
which can exhibit a phase transition between a fluid and a condensed
phase,  
similar to behaviour  encountered in models of random geometries
in one, two and four dimensions.  
This model can be viewed as a generalisation of the backgammon model introduced
by Ritort as an example of glassy behaviour without disorder. \vspace{5mm}\\
{ITFA-96-38, BI-TP 96/47}
\end{abstract}

\end{frontmatter}

\section*{Introduction}

Given the success of the dynamical triangulation approach to
simulating two dimensional quantum gravity \cite{1}, there has been a
concerted effort to employ similar methods in understanding higher
dimensional discretized gravity models \cite{hist}.  In three and four
dimensions one does not have the underpinning of matrix models to give
confidence in the simulation results, but the general phase diagram of
such models is now fairly clear.

One finds a two-phase structure with 
an elongated/branched polymer phase and a crumpled/collapsed phase.
Recent results suggest strongly
that the transition, originally thought to be
continuous, is first order \cite{first}.
In addition, the vertex order distribution in the crumpled
phase displays some peculiarities \cite{hin,ckrt}. In D-dimensional models
a singular $D-2$ simplex appears in the crumpled phase and mops
up an extensive fraction of the volume. 

A similar phase structure is also encountered in one and two dimensional
systems \cite{adfo,bb,aj1} but only recently was it 
hypothesized that the
same mechanism might be responsible in all the cases \cite{bbpt}.  
The
suggestion in \cite{bbpt}
was that an effective polymer model described not only the
elongated phase but also the essential features
of the transition and the crumpled phase.  In \cite{bbpt} it was
pointed out that the polymer model admitted a description in terms of
balls in boxes, with a box being a vertex and the number of balls in a
box being the vertex order.  The balls in boxes metaphor has already
occurred in the analysis of glassy dynamics \cite{r}, where the name
``Backgammon models'' was coined.

In this paper we analyse the density driven transition of \cite{bbpt}
using the methods of \cite{r}. We first discuss the balls--in--boxes
model in general terms, before going on to describe the transition
itself. In closing we highlight the similarities with Bose-Einstein
\cite{BE,Feyn} condensation and the treatment of the spherical model
phase transition in the classical paper of Berlin and Kac \cite{Kac}.
The mapping between the balls--in--boxes model and the BP model of
\cite{bb} is made completely explicit in an appendix.

\section*{The balls--in--boxes model}

Consider the system of $N$ positive integer numbers: $q_i\quad i=1,\ldots,N$.  
We define the partition function:
\begin{eqnarray}
Z(N,M)=\sum_{q_1,\ldots,q_N}p(q_1)\cdots p(q_N)
\delta_{\textstyle q_1+\cdots+q_N,M}.
\label{Z}
\end{eqnarray}
The $p(q)$ is some given function on positive integers taking on
non-negative real values. If it is interpreted as a probability we
have the additional requirement that $\sum_q p(q)=1$.  We can view
this system as an ensemble of $M$ balls distributed into $N$
boxes. Each box is weighted by the factor depending only on the number
of balls it contains.  
In the glassy dynamics applications the choice in \cite{r} 
was $p(q) = \exp
( \beta \delta_{q,0})$ and there was no static phase transition. The
natural choice of $p(q)$ for models associated with random geometries
is $p(q) \sim q^{-\beta}$, as such factors appear as discretizations
of terms coming from the metric integration measure.  In what follows
we will keep our discussion as general as possible, specializing to
$p(q) \sim q^{-\beta}$ where appropriate.

We are interested in the system with a fixed density, {\em i.e.} the
average number of balls per box, $\rho=M/N$.
Inserting the integral representation for the delta function into
(\ref{Z}) we obtain:
\begin{eqnarray}	
\lefteqn{Z(N,\rho)=\frac{1}{2\pi}\sum_{q_1,\ldots,q_N}p(q_1)\cdots p(q_N)
\int\limits_{-\pi}^{\pi}\!\mbox{d}\lambda 
e^{-i\lambda(q_1+\cdots+q_N-\rho N)}=}\nonumber\\
&&\frac{1}{2\pi}\int\limits_{-\pi}^{\pi}\!\mbox{d}\lambda e^{i\lambda \rho N}
\left(\sum_{q}p(q)e^{-i\lambda q}\right)^N
\label{Zint}
\!=\frac{1}{2\pi}\int\limits_{-\pi}^{\pi}\!\mbox{d}\lambda 
\exp\left(N (i\lambda\rho+\log F(e^{-i\lambda})\right)
\end{eqnarray}
where $F(z)=\sum_{q=1}^\infty p(q)z^q$. 

The generating function $F(z)$ contains  all the information about the
system. In this paper we will consider only the case when  this
function has a finite, non-zero radius of convergence $z_c$.  Notice
that without the loss of generality we can put $z_c=1$, since due to
the constraint $q_1+\dots + q_N = M$ the change $p(q) \rightarrow p(q)
z_c^{-q}$ introduces only a constant factor $z_c^{-M}$ in front of the
sum defining the partition function (\ref{Z}). From here on we keep
$z_c=1$.  We further consider two subcases~: {\bf (a)} the derivative
of $F(z)$ at the radius of convergence: $F'(1) = \sum_q q p(q)$ is
finite, and {\bf (b)} $F'(1)$ is infinite.

The integral (\ref{Zint}) can be calculated by the saddle point method.
By deforming the contour in the lower half plane  
$\mbox{Im}\, \lambda < 0$ 
we find that the saddle point is the solution of the equation~:
\begin{eqnarray}\label{sp} 
\rho  = -\frac{d}{d \mu} \log F(e^{-\mu}) = G(\mu) 
\end{eqnarray}
and lies on the imaginary halfaxis $\lambda_0 = -i\mu$ 
for a certain range of $\rho$.
To see this we notice that the function $G(\mu)$ increases
monotonically between the values $G(\infty)=1$ to $G(0) = F'(1)/F(1)$ 
when the argument $\mu$ goes from infinity to zero. As we will
see $G(0)$ corresponds to the critical density and we 
denote it by $\rho_c$. Since the function $G(\mu)$ is monotonic 
the equation (\ref{sp}) can be solved for real positive 
$\mu = \mu(\rho)$ with $\rho$ between $1$ and $\rho_c$. 
The leading part of the free energy density $f(\rho)$ calculated
by the saddle point method is then 
\begin{eqnarray} 
f(\rho) = \frac{1}{N} \log Z(N,\rho) = \rho \mu + \log F(e^{-\mu})
\label{fe}
\end{eqnarray}
and grows from zero to 
\begin{eqnarray} 
f(\rho) = \frac{1}{N} \log Z(N,\rho) = \log F(1)
\label{fsing}
\end{eqnarray}
when the density grows from one to $\rho_c$.
This value of the free energy is in fact the upper limit for the 
free energy density and corresponds to the situation when all, or almost
all, the probabilities $p(q_i)$ are decoupled~:
\begin{eqnarray}
Z(N,M)=\sum_{q_1,\ldots,q_N}p(q_1)\cdots p(q_N) =
\left(\sum_q p(q)\right)^N = e^{ \displaystyle N \log F(1)}.
\end{eqnarray}
When $\rho$ is larger than $\rho_c$ the free energy is blocked at the
limiting value $\log F(1)$ and does not change further with $\rho$.
Thus, there are two phases: for $\rho<\rho_c$ the free energy in the
thermodynamic limit is given by (\ref{sp}) and (\ref{fe}), while for
$\rho> \rho_c$ the free energy is given by (\ref{fsing}). To
distinguish those two phases one can define the following expression
\begin{eqnarray}\label{op}
\frac{\partial f}{\partial\rho}= \mu(\rho)
\end{eqnarray}
which vanishes above $\rho_c$ and can serve 
as an order parameter for this system. 
Notice that the terms proportional to the derivative $d \mu / d\rho$ 
arising from differentiating (\ref{fe})  
vanish due to the saddle point equation (\ref{sp}). 

We will see later that for $\rho>\rho_c$ the finite size corrections
to the formula (\ref{fsing}) are of the form $\log p((\rho-\rho_c)N)$ and
come from the energy of a singular box containing an extensive
$(\rho-\rho_c)N$ number of balls. This is the phenomenon of condensation,
and we call this phase the  {\em condensed} phase. The other phase, for
$\rho < \rho_c$, we call the  {\em fluid} phase.

In case {\bf (b)} the situation is different, since 
$F'(1)/F(1)$ is infinite, and therefore the critical density $\rho_c$
is also infinite. This means that the system never enters the 
condensed phase and has only the fluid one. Similarly one can check that
the model has no condensed phase when the radius 
of convergence of $F(z)$ is infinite. 

The expression for the free energy is 
identical to those obtained in the branched polymer (BP) model
\cite{bb} which has $\rho=2$. We show explicitly in the
Appendix that the BP model can be
mapped directly 
onto the balls--in--boxes model.

\section*{Critical exponents}

In this section we will calculate critical exponents for
the specific case~: $p(q) \sim q^{-\beta}$.
As we have already noted this is the natural weight
to consider for models associated with random geometries
(and is that used in the BP model of \cite{bb}).
These weights are also known to appear in the broad class of models
with self organized criticality (see for example \cite{Sorin} and references within). Adding an exponential
prefactor to the weights $p(q) \sim e^{\kappa q} q^{-\beta}$ will
not affect the phase structure of the model, since as discussed before,
it results only in the factor $e^{-\kappa M}$ in front of  the partition
function $Z(N,M)$.

The generating function is
thus given by the Dirichlet series \cite{apos} 
$F(z) = \sum_{q=1}^{\infty} q^{-\beta} z^q$, and the critical
density is given by:
\begin{eqnarray} 
\rho_c= F'(1)/F(1) = \frac{\zeta(\beta-1)}{\zeta(\beta)}
\end{eqnarray}
where $\zeta(\beta)$ is the Riemann Zeta function. For $\beta>2$
the critical density is finite. It diverges when $\beta$ approaches two
and the system never enters the condensed phase. 
For $\beta$ smaller than two the model has only the fluid phase.
When the density approaches the critical one 
$\Delta \rho = \rho_c-\rho \rightarrow 0^+$, then the free energy density 
becomes singular. The first derivative of the free energy density with respect
to $\rho$ is equal $\mu(\rho)$ which is singular. To see this let us
expand the function $G(\mu)$ on the right hand side of (\ref{sp}) 
for small $\mu$. There are two cases. For 
$2<\beta<3$ the expansion starts from the singular term \cite{bb}~:
\begin{eqnarray}
\Delta \rho = G(\mu) = b \mu^{\beta-2} + \dots
\end{eqnarray}
where $b$ is a positive constant depending on $\beta$. 
For $\beta>3$ the dominating part
comes from the analytic part of $G(\mu)$~:
\begin{eqnarray}
\Delta \rho = G(\mu) = a \mu + \dots + b \mu^{\beta-2} + \dots
\end{eqnarray}
where $a$ is also a constant depending on $\beta$. The first dots
denote other powers of $\mu^n$ coming from the analytic part for
$n<\beta-2$. Inverting the last two equations 
to solve for $\mu$ gives,
in the limit $\Delta \rho \rightarrow 0^{+}$~:
\begin{eqnarray}
\frac{\partial f}{\partial\rho} = \mu \sim \left\{
\begin{array}{ll}
\Delta\rho^{1/(\beta-2)} & \quad \mbox{ for } 2 < \beta < 3 \\
\Delta\rho^{(\beta-2)} & \quad \mbox{ for } \beta > 3
\end{array}
\right.
\end{eqnarray}
For integer $\beta$ one has logarithmic corrections, in particular for $\beta=3$,
$\mu = \Delta \rho \log \Delta \rho$. By similar reasoning one
can find the  singularities of the free energy with respect to
the parameter $\beta$ \cite{bb}. 

\section*{Dressed box-occupation distribution}

If it were not for the constraint the probability of finding 
$q$ balls in one box would be equal to the {\em bare} probability 
$p(q)$. Because of the constraint the 
relation between the {\em effective} or {\em dressed}
probability $\pi(q)$ and the bare probability
$p(q)$ is given by:
\begin{eqnarray} 
\pi(q)&=&\frac{p(q)}{Z(N,M)} \sum_{q_2\ldots,q_{N}} 
p(q_2)\cdots p(q_{N})
\delta_{\textstyle q+q_2+ \cdots+q_N,M}\nonumber\\
&=&
p(q)\frac{Z(N-1,M-q)}{Z(N,M)}
\end{eqnarray}
The above expression is simply derived
by observing that when $q$ is 
fixed in one box the other $N-1$ boxes must contain $M-q$
balls. In the thermodynamic limit $\rho = const$ and
$N\rightarrow\infty$ we can use the saddle point result
\begin{eqnarray} 
Z(N,M) \sim \exp\left( \mu M +\log F(e^{-\mu}) N\right)
\end{eqnarray}
to obtain:
\begin{eqnarray} 
\pi(q)=\frac{p(q) e^{-\mu q}}{F(e^{-\mu})}
\label{pkl}
\end{eqnarray}
in the fluid phase and 
\begin{eqnarray}
\pi(q)= \frac{p(q)}{F(1)}
\label{pks}
\end{eqnarray}
in the condensed phase. 

We can now repeat the same reasoning  for the two (or multi)--point
probabilities. We fix the occupation numbers $q_a$ and $q_b$
in two boxes and find that the dressed two--point probability
is~:
\begin{eqnarray} 
\pi(q_a,q_b) = p(q_a)p(q_b) \frac{Z(N-2,M-q_a-q_b)}{Z(N,M)}
\end{eqnarray}
Calculating the last expression in the thermodynamic
limit and comparing it with the one point probabilities
and finds that in this limit occupation numbers in different 
boxes become independent, namely~:
\begin{eqnarray} 
\pi(q_a,q_b) - \pi(q_a) \pi(q_b) \rightarrow 0 
\end{eqnarray}
and the same holds for other the higher multi--point 
probabilities.

For every finite N the distribution
$\pi(q)$ must satisfy the constraint:
\begin{eqnarray} \label{constr}
\sum_q q\pi(q)=\rho.
\end{eqnarray}
It is easy to check that this constraint is satisfied by the the
limiting distribution (\ref{pkl}) in the fluid phase and at the
transition.  Exactly at the transition where $\mu=0$ the constraint
is satisfied by the bare distribution $p(q)$. 
As we move into the phase condensed phase
$\rho > \rho_c$, the constraint is not  satisfied by (\ref{pks}),
which implies that the limit is not uniform. 

In the fluid phase the the constraint (\ref{constr}) can be enforced
by the exponential term in (\ref{pkl}). In the condensed phase
this would require $e^{-\mu}=z>1$ 
which is impossible because of singularity.
The probability (\ref{pks}) stays equal  to the bare probability
and the system compensates otherwise for the ``missing'' particles.
This change in behaviour produces a phase transition,
and is perhaps best understood by looking at the
behaviour of $\pi(q)$ for a finite number of boxes.

\section*{Finite number of boxes}

The method described in the previous sections works only in the
thermodynamical limit. As pointed out already the limiting distribution
(\ref{pks}) does not satisfy the constraint and indicates some substantial
finite size effects. To study those effects we used two methods:
\begin{enumerate}
\item Exact calculation of partition function.\\
The partition function (\ref{Z}) satisfies the following recursion relation:
\begin{eqnarray}\label{relation}
Z(N,M)=\sum_{q=1}^M Z(N-1,M-q)p(q),\qquad N\ge2
\end{eqnarray}
with initial conditions $Z(1,q)=p(q)$.\\
Equation (\ref{relation}) can be used to calculate the partition function
and in principle any of its derivatives. 
The cost of calculating Z is proportional to $NM^2$ but
the actual limitation on the size of the system which one can study
using this method is given by the numerical precision
as the value of $Z(N,M)$ drops exponentially with $N$. 
We have found that the practical limit is about $N=500$. 
\item Monte-Carlo simulations.\\ 
For the bigger systems we are left so
far with the necessity of MC simulations.  The construction of the
appropriate program is standard and we will not describe it here. The
program allows to simulate the system of few thousands boxes with few
hours of CPU time on an average workstation.
\end{enumerate}
\begin{figure}[t]
\begin{center}
\epsfig{file=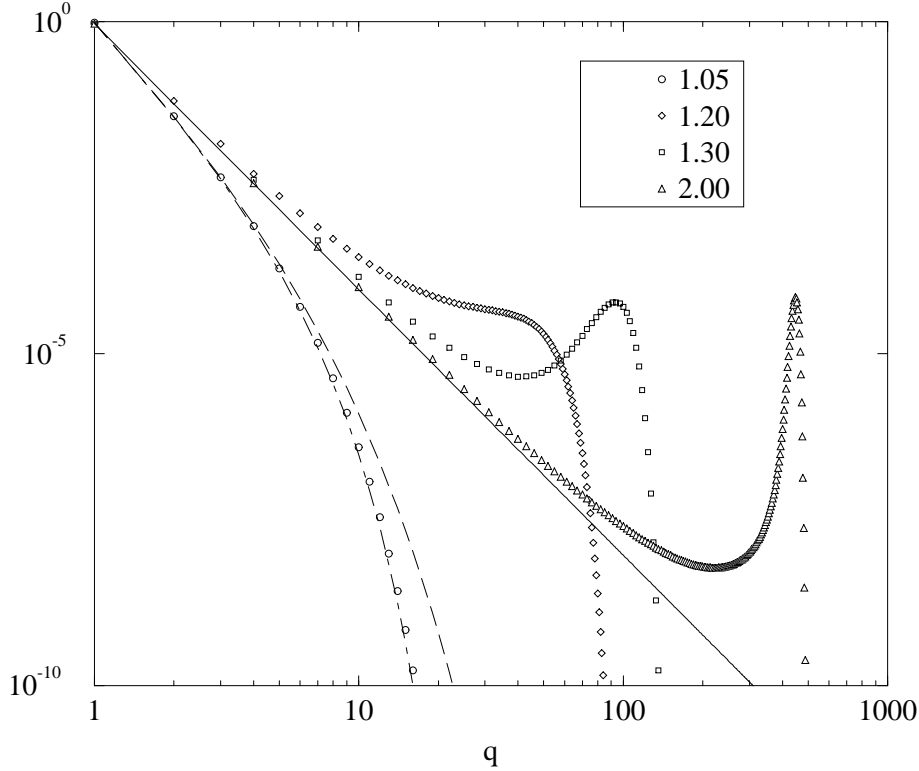,width=14cm,bbllx=18,bblly=60,bburx=552,bbury=430}
\end{center}
\caption{\label{frho}The effective one--point distributions for
different $\rho$ and  500 boxes. The solid line is the bare distribution 
$p(q)\sim q^{-4}$. 
The finite-size corrected $\pi(q)$ distribution for $\rho=1.05$ 
is given by the dot-dashed line, while the long dashed line is the thermodynamical limit.}
\end{figure}

Using both these methods one confirms the following picture 
of the system:
In the fluid phase the distribution nicely converges
to the limiting distribution (\ref{pkl}) as expected.
In the condensed phase the  distribution follows
the limiting power-like distribution (\ref{pks}) 
for smaller $q$ 
but develops a peak in the tail.
These effects are clearly seen on the 
figure~\ref{frho} which illustrates
the phase transition produced by change in balls density. 

One can include the finite size effects into the formula
(\ref{pkl}) by noting that  $\mu$ depends on $\rho$
and hence on $q$ since $\rho=(M-q)/N$. The predicted distribution is plotted
on the figure~\ref{frho} (dashed line) and the agreement with 
the exact  result is excellent. The same reasoning can be applied to the formula (\ref{pks}). This results in sharp decrease of distribution at large $q$,
clearly indicating a different nature of finite size effects in this case.

\begin{figure}[t]
\begin{center}
\epsfig{file=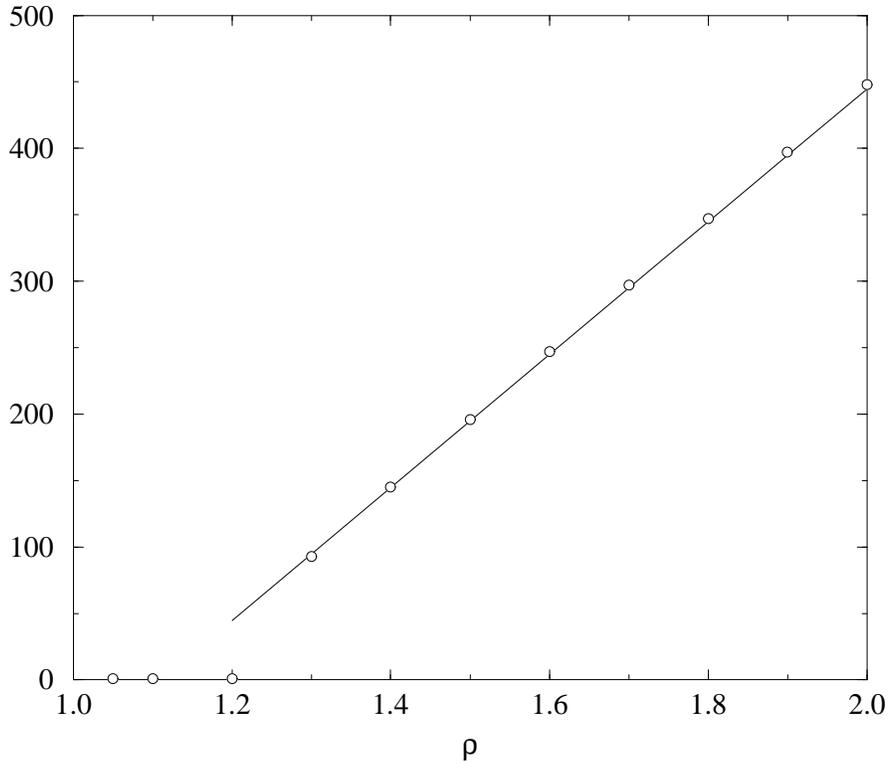,width=14cm,bbllx=18,bblly=60,bburx=552,bbury=430}
\end{center}
\caption{\label{fpeak}Position of peak as the function of $\rho$.
($\beta=4.0$, 500 boxes)}
\end{figure}

The peak in the tail appears as the system tries to compensate for the
particles missing from (\ref{constr}).  We can approximately predict
the position of this peak by the following reasoning: The number of
missing particles is
\begin{eqnarray}\label{peak} 
N(\rho-\sum_q q \pi(q))=
N\left(\rho-\frac{F'(1)}{F(1)}
\right)
=N(\rho-\rho_c)
\end{eqnarray}
so assuming  that all the missing particle are in one box
this is  the position of the peak, which as moves linearly with increasing $N$.
In figure~\ref{fpeak} we plot the position of the peak
compared with  prediction (\ref{peak}).
As one can see, the agreement is excellent.

\section*{Discussion}

The density driven transition obviously has
a strong similarity to Bose-Einstein condensation \cite{BE}, especially
as presented in, for instance, \cite{Feyn}. Although this is usually
formulated in a grand canonical ensemble with a chemical potential
the discussion may be paraphrased into language very similar to that used
above.
If one fixes the number of energy levels to $N$ 
and the number of particles to $M = \rho N$ the partition function 
for Bose particles is
\begin{equation}
Z = \sum_{q_1,\ldots,q_N} \prod_{i=1}^N \exp( -\beta q_i \epsilon_i ) \; \delta(\sum_{i=1}^N q_i -  M)
\end{equation}
where each box has its own energy level $\epsilon_i$.
If one slips into a continuum language
the expression one arrives at for the particle density 
(in three dimensions) is just
\begin{equation}
\tilde \rho = s \left( { m k T \over 2 \pi h^2} \right)^{{3 \over 2}} 
F(z)
\end{equation}
where $\tilde \rho$
is now the particle density per unit volume,
$s$ is a spin degeneracy factor and $F(z)= \sum_{q=1}^{\infty}
q^{-3/2} z^q$.
At fixed particle density 
per unit volume $z$ is pushed to one as $T$ is reduced, 
at which point
a sizeable proportion of the particles begin to 
condense in the ground state. 
Thus at the formal level the mechanism for the
Bose-Einstein condensation and the transition 
discussed in this paper are identical, namely,
a Dirichlet series is pushed to the limit of its radius of convergence.
The similarities with the treatment of the spherical model
phase transition in \cite{Kac} are also apparent. In that case too
the (spherical) constraint is exponentiated and a saddle point
``sticking'' signals the phase transition.

We would like to stress, however, that in spite of formal similarity
there exists a substantial  difference. In the Bose-Einstein condensation 
case each box has its own energy level and the energy is 
linear in number of balls. The system condenses into
the lowest state. In the Backgammon case the 
boxes are equivalent and condensation is a kind of symmetry breaking process.
Moreover, the energy of the box is {\em not} linear in number of balls 
which is absolutely crucial for existence of phase transition,
so there is no direct interpretation in terms of occupied states. 
One way of looking at the model would be to imagine 
a set of $N$ ``atoms'' and $M$ ``electrons''. The electrons would be bound to
the atoms and the energy of an atom with $n$ electrons would be $\log(n)$,
corresponding to the energy levels $E(k)=1/k$. 

Variations of the model with different vertex weights can also
be easily accommodated into the saddle point approach. For instance,
changing the weight of a vertex of order $q$ 
from $1/q^{\beta}$ to $t_q /q^{\beta}$
gives the following equation for the
critical density
\begin{equation}
\rho_c = { \zeta(\beta-1) + q \left( {t_q - 1 \over q^{\beta}}\right)
\over
\zeta(\beta) +  \left( {t_q - 1 \over q^{\beta}}\right) }.
\end{equation}
For univalent vertices we can see
from the above that as $t_1 \rightarrow 0$ the asymptote for $\rho_c$
moves from one to two, so we reproduce the result of \cite{bb} that the
transition is suppressed for finite $\beta$ when $\rho=2$.

In the case of infinite radius of convergence of $F(z)$ the condensation
does not take place.  Clearly, this will happen when we restrict the
number of balls that are allowed in one box ($p(q)=0,\;
q>q_{\mbox{max}}$) as the function $F(z)$ is then a polynomial in
$z$.

The appearance of the $\zeta$ function in the saddle point equation
also encourages one to play number-theoretic games with the vertex
weights. It is possible to regard the standard $\zeta$ function as a
bosonic partition function and define ``arithmetic'' k-parafermions
\cite{bow} with the use of a generalized M\"obius function
\begin{equation}
\mu_k (q) = 1 \; \; \mbox{if} \; \; q=\prod_i (p_i)^{r_i}, \; 0 \le r_i \le k-1
\end{equation}
where the $p_i$ are the prime factors of $q$.
The case $k=2$ corresponds to arithmetic fermions
and $k=\infty$ to bosons (the standard $\zeta$ function).
If the vertex weights are taken to be
$\mu_k (q) / q^{\beta}$ the
saddle point equation becomes
\begin{equation}
\rho = \left[ { \zeta(\beta-1) \over \zeta(\beta)}
- { \zeta(k \beta-1) \over \zeta(k \beta)} \right].
\end{equation}
The sort of fermionic partition functions generated here
are thus not useful for suppressing a collapse transition.
The vertex weights do not suppress large numbers of particles
in a box, only those numbers with $k$ identical prime factors
are not allowed. The behaviour of $\rho_c$ as a function of
$\beta$ is thus not much changed, apart from the asymptote
for large $\beta$ being reduced to zero. These arithmetic
parafermion models thus crumple at a {\em lower}
$\rho_c$ for a given $\beta$.

The backgammon models are perhaps
best thought of as models for the transition 
in 4D gravity at the level 
of baby universes and their connecting ``necks''
\cite{bbpt}, identifying the polymer structure
with the baby universe tree.
As was noted in \cite{bbpt}, the 4D simulations have
a fluctuating density, as (almost) fixing $N_4$ still allows
$N_0$ and hence $\rho$ to vary, but such a smearing
of the density should
not affect the saddle point equations and the resulting
phase structure. The appearance of models
with $p(q) \sim q^{-\beta}$ is not, as we have noted above,
restricted to those arising in the context of discretized gravity simulations.
In the  model described in \cite{Sorin}
the power exponent is generated dynamically and
is calculated by assuming that the distribution function
is properly normalised. This corresponds exactly to our 
constraint (\ref{constr}). This constraint is satisfied by a pure
power law distribution only at the transition.

The backgammon models of \cite{r} 
were originally formulated to investigate glassy
behaviour, and their dynamics display many spin-glass like features.
Although the energy landscape of the polymer
inspired backgammon models here is a long way from that
of the ``golf-course'' landscape of the original backgammon models it
might still be a worthwhile exercise
to investigate the dynamics of the polymer backgammon models
in the light of the results of \cite{r}. The dynamics may prove
as interesting as the static properties discussed in this paper.

\section*{Acknowledgements}
We are grateful to S.~Bilke, B.~Petersson, J.~Tabaczek, F.~Ritort and J.~Smit
for many valuable discussions.
P.B. thanks the Stichting voor Fundamenteel Onderzoek der Materie
(FOM) for financial support. Z.B. has benefited from the financial
support of the Deutsche Forschungsgemeinschaft under the contract Pe
340/3-3. This work was partially supported by KBN (grant 2P03B 196
02) and by EC HCM network grant CHRX-CT-930343.

\section*{Appendix A}

In this appendix we describe in detail the mapping between
the branched polymer model of \cite{bb} and the
balls--in--boxes model.

Let $n_q$ be the number of boxes with $q$ balls in it. 
These numbers must satisfy the constraints:
\begin{eqnarray}\label{nconstr}
\sum_qn_q=N,\qquad\sum_q q n_q=M
\end{eqnarray}
The number of configurations with given $n_q$'s 
is given by (we assume $0!=1$)
\begin{eqnarray}\label{nn}
\frac{N!}{n_0!n_1!\cdots n_N!}
\end{eqnarray}
so the partition function (\ref{Z}) can be written as:
\begin{eqnarray}\label{ZZ}
Z(N,M)=\sum_{\{n_i\}}\frac{N!}{n_0!n_1!\cdots n_N!}p_0^{n_0}p_1^{n_1}p_2^{n_2}\cdots p(n)^{n_N}
\end{eqnarray}
where the sum runs over all sequences of positive integers
satisfying (\ref{nconstr}).

The number of rooted, planted, planar trees with 
$v_1$ vertices of degree one, $v_2$ vertices of degree two, and so on
is given by (see for example \cite{comb}):
\begin{eqnarray}\label{vv}
\frac{1}{N}\frac{N!}{v_1!\cdots v_N!}
\end{eqnarray}   
These numbers satisfy
\begin{eqnarray}
\sum_qv_q=N,\qquad\sum_q q v_q=2N-1
\end{eqnarray}
We see that
the weight of a branch polymer depends on the vertex orders in the same
way as a balls--in--boxes configuration 
depend on the box occupation numbers\cite{adfo,bb}.
Therefore the BP partition function can be also written in 
the form (\ref{ZZ})
with the factor (\ref{nn}) substituted by (\ref{vv}). 
If we now 
forbid the empty boxes by putting $p(0)=0$ and set $M=2N-1$ ($\rho=2$) those
two functions will be equivalent.

\end{document}